\documentstyle[aps,amssymb,graphicx,twocolumn]{revtex}
\DeclareSymbolFont{rsfscript}{OMS}{rsfs}{m}{n}
\DeclareSymbolFontAlphabet{\mathrsfs}{rsfscript}

\newcommand{\nnn}{\nonumber\\}

\newcommand{\eref}[1]{(\ref{#1})}

\newcommand{\rmd}{{\rm d}}

\newcommand{\bra}[1]{\langle#1|}
\newcommand{\ket}[1]{|#1\rangle}

\newcommand{\Len}{{\mathrsfs{L}}}  %
\newcommand{\Area}{{\mathrsfs{A}}} %

\newcommand{\tfrac}[2]{{\textstyle\frac{#1}{#2}}}

\newcommand{\Natural}{{{\mathbb{N}}}}

\newcommand{\PO}{\ .}   %
\newcommand{\CO}{\ ,}   %
\newcommand{\oh}{{\frac{1}{2}}} %

\newcommand{\rvec}{{\bf r}}

\newcommand{\vvec}{{\bf v}}

\newcommand{\wasn}{i} %
\newcommand{\mm}{{\rm m}} %
\newcommand{\mmsm}{\overline{\rm m}} %

\newcommand{\dedge}{d_{\rm edge}}
\newcommand{\dedgesm}{\overline{d}_{\rm edge}}
\newcommand{\dosc}{\tilde{d}}
\newcommand{\N}{{\rm N}}

\newcommand{\Nsm}{\overline{\N}}

\newcommand{\po}{{\gamma}}

\title{Spectral cross correlations of magnetic edge states}

\author{Klaus Hornberger$^{1,2}$ and Uzy Smilansky$^2$}
\address{$^1$Max-Planck-Institut f\"ur Physik komplexer Systeme,
01187 Dresden,  Germany}
\address{$^2$The Weizmann Institute of Science, 76100 Rehovot, Israel}

\date{To be published in Phys. Rev. Lett. {\bf 88}
  024101  (2002)}

\begin{document}
\maketitle
\begin{abstract}
  We observe strong, non-trivial cross-correlations between the edge
  states found in the {interior} and the {exterior} of magnetic
  quantum billiards.  
  Our analysis is based on a novel definition of the edge state spectral
  density which is rigorous, practical and semiclassically accessible.
\end{abstract}
One of the main goals in the field of ``quantum chaos'' is to link the
auto-correlations found in a quantum spectrum to the periodic
orbits of the classical problem 
\cite{Gutzwiller98}.
Here, we extend this study and investigate whether {\em
  cross}-correlations exist between quantum systems which are
different but related by their classical dynamics. We develop this
idea for magnetic quantum billiards
\cite{NP67,B:Y:J:95,BB97aNE} which often
serve to model
semiconductor quantum dots 
\cite{URJ95}.

Magnetic billiards consist of a charged particle moving
ballistically in a compact domain in the plane subject
to a 
homogeneous magnetic field.
The quantum wave function is required to vanish at the billiard
boundary  while the impinging classical particle is
reflected specularly \cite{RB85,BK96,Gutkin01}.
The boundary defines also a complementary problem -- an anti-dot --
where the particle is confined to the exterior, and is scattered at
the billiard boundary.  Although the exterior domain is unbounded, its
spectrum is {\em discrete} \cite{HS00a}.
Is it possible to relate the energy levels of the dot to those of
the anti-dot?
We shall show that  there exists an intimate, non-trivial
connection between the spectra of the interior and the exterior
problem.
It is the quantum manifestation of a duality in the classical dynamics.

\begin{figure}[b]%
  \begin{center}%
    \includegraphics[width=\linewidth]{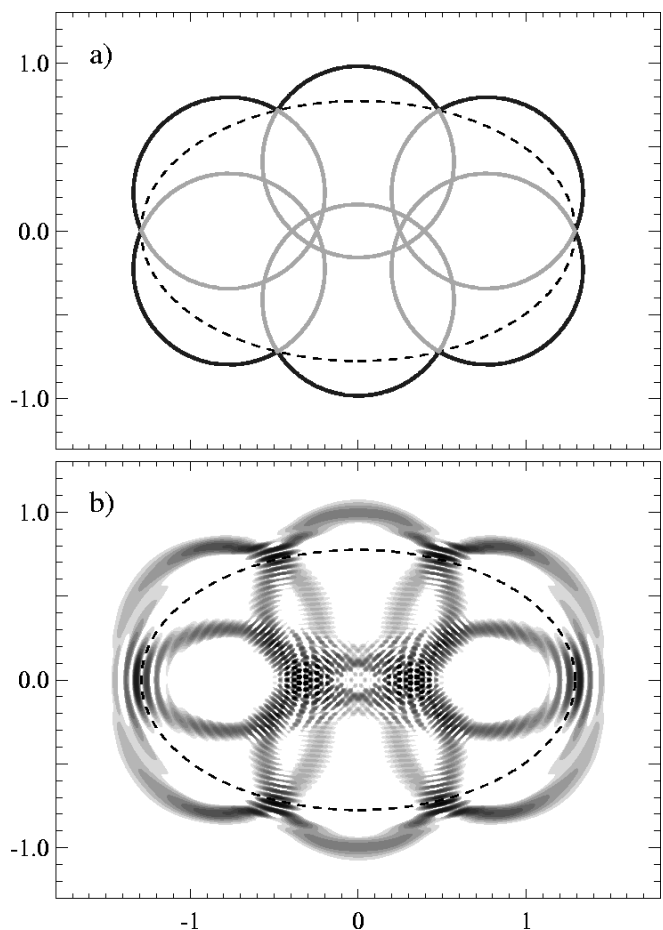}
  \end{center}%
  \caption{(a) A pair of dual periodic trajectories and (b)
    a pair of correlated eigenfunctions found in the interior
    and the exterior ellipse billiard (superimposed). The
    billiard boundary is indicated by a dashed line.}
  \label{fig:cpair}
\end{figure}

The classical {\em interior-exterior duality} is illustrated in
Fig.~\ref{fig:cpair}(a): Since a (periodic) orbit consists of arcs of
constant curvature, one can construct a {\em dual} orbit in the
complementary domain by completing the arcs to circles.
Any skipping trajectory meets with a dual one under rather
general conditions -- if every circle of cyclotron radius $\rho$
intersects the boundary at most twice.
Pairs of dual periodic orbits have the same stability and their
actions add up to an integer multiple of the action of a cyclotron
orbit.  On semiclassical grounds one may therefore expect the
correlation between the interior and the exterior motion to carry over
to the quantum spectrum.
This is also corroborated by the existence of
pairs of interior and exterior quantum eigenstates which
match up well,  cf  Fig.~\ref{fig:cpair}(b),
although their energies differ.

The semiclassical analysis is complicated by the fact that in the
exterior each Landau level is an accumulation point for an infinite
series of energies.  The respective states -- the {\em bulk states} --
correspond to unperturbed cyclotron motion.
Also in the interior one may find (a finite number of) bulk states
if $\rho$ permits
complete cyclotron orbits to fit into the domain.
The eigenfunctions
which correspond to the skipping trajectories, on the other hand,
are called {\em edge states}.  Clearly, a possible correlation is to
be expected only between these non-trivial exterior and interior states.

Although the notion of edge states is intuitively clear and often used
(e.g. in the context of  the quantum Hall effect \cite{Halperin82}),
we are not aware of a general quantitative definition in the
literature (however see \cite{AANS98}).  Therefore, the purpose of
this letter is twofold. First, we propose a definition for the
spectral density of edge states which provides a meaningful
characterization and applies
in the quantum and in the semiclassical regime.
Only with this can we then establish the existence of a pairwise
relation between the edge states of the interior and the exterior.

A definition for edge states should take into account  that 
a clear separation into
edge and bulk occurs only in the semiclassical limit $b\to 0$.  Here,
we  express the quantum scale in terms of
the magnetic length
$b=\sqrt{2\hbar/(m\omega_{\rm c})}$ (with $m$ the mass and
$\omega_{\rm c}$ the cyclotron frequency).
At finite values of $b$
the states  may be of
an intermediate type.  We
propose to quantify the transition by attributing a weight $w_\wasn>0$ to
each eigenstate ${\psi_\wasn}$ (of energy $\nu_\wasn$) which gives a
measure of the degree to which ${\psi_\wasn}$ has the character of an
edge state.  The spectral density of edge states in either the
interior or the exterior is then defined as
\begin{equation}
  \label{eq:dedgedef}
  \dedge(\nu)=
  \sum_{\wasn=1}^\infty w_\wasn\, \delta(\nu-\nu_\wasn)
  \PO
\end{equation}
Here, we scale the energy $E$ by the spacing between Landau levels,
$\nu={E}/{(\hbar\omega_{\rm c})} = {\rho^2}/{b^2}$.
Any reasonable definition of the weights $w_\wasn$ must suppress the
bulk states by exponentially small values,
such that the mean edge density $\dedgesm$ is well-defined in the
exterior and equal to the interior one, to leading order.  In the
semiclassical limit it should match our notion of edge states
admitting a trace formula which involves only the skipping
trajectories.
Moreover, we shall demand $\dedgesm$ to coincide with the unweighted
interior mean density if the cyclotron radius is large
enough to prevent bulk states.

To motivate our definition of the weights $w_\wasn$, consider  the {\em
scaled} magnetization
${\mathcal{M}}$ of the interior billiard, a sum over the scaled
magnetic moments

\begin{equation}
  \label{eq:Magdef}
  {\mathcal{M}}(\nu;b)
  =\sum_{\nu_\wasn\leq\nu}
  \frac{ \bra{\psi_\wasn}\rvec\times\vvec\ket{\psi_\wasn}}{\omega_{\rm c}b^2}
  = \int_0^\nu \!\mm(\nu';b) \, \rmd\nu'
\PO
\end{equation}
The scaled magnetization density $\mm(\nu)$  can be expressed 
by the
derivatives of the spectral counting function
$\N(\nu;b)=\sum_\wasn\Theta(\nu-\nu_\wasn(b))$ with respect to $b^2$ and
$\nu$,
\begin{equation}
  \label{eq:mmdef}
  \mm(\nu)=-b^2\frac{\partial \N}{\partial b^2}
 - \nu \frac{\partial  \N}{\partial \nu}
= \sum_\wasn \Big( b^2\frac{\rmd \nu_\wasn}{\rmd b^2}-\nu_\wasn\Big)
\,  \delta(\nu-\nu_\wasn)
\PO
\end{equation}
This is verified by replacing the energies $\nu_\wasn$ in
\eref{eq:mmdef} by the expectation values of the Hamiltonian.  The
scaled magnetization density may be obtained from the
conventional one by multiplication with the field strength $B$. 
It exhibits a natural partitioning into a bulk part and an edge part 
since it complies with the scaling properties of the system:
The scaled magnetic moment of a Landau state is $-\nu$. Hence,
the second part of \eref{eq:mmdef},
\begin{eqnarray}
  \mm_{\rm bulk}(\nu) &=&
   - \nu \frac{\partial  \N}{\partial \nu} =
  \sum_\wasn (-\nu_\wasn)\, \delta(\nu-\nu_\wasn)
\CO
\end{eqnarray}
attributes the full
diamagnetic response of a Landau state
to each state $\psi_\wasn$.  We call it the bulk magnetization
density.  It follows that the remaining part, the {\em edge magnetization}
density
\begin{eqnarray}
  \label{eq:mmedgedef}
  \mm_{\rm edge}(\nu) =
  - b^2 \frac{\partial\N}{\partial b^2}=
 \sum_{\wasn=1}^\infty b^2\frac{\rmd \nu_\wasn}{\rmd b^2}
  \,\delta(\nu-\nu_\wasn)
\end{eqnarray}
assigns the positive {excess} magnetic moments induced by the
presence of a billiard boundary. Its mean value
\begin{equation}
  \label{eq:mmedgesm}
  {\mmsm}_{\rm edge}(\nu)
  =  \frac{\Area}{b^2\pi} \nu
  - \frac{1}{2}\,\frac{\Len}{2\pi b} \nu^\frac{1}{2}
  = - {\mmsm}_{\rm bulk}(\nu)
\end{equation}
follows  from the mean number of states in a magnetic
billiard with area $\Area$ and circumference $\Len$
 \cite{MMP97},
\begin{equation}
  \label{eq:Nsmooth}
  \Nsm(\nu;b)
  = \frac{\Area}{b^2\pi} \, \nu
  - \frac{\Len}{2\pi b} \, \nu^\oh  + \frac{1}{6}
\PO
\end{equation}
Note that $ {\mmsm}_{\rm edge}$ cancels the mean bulk magnetization
density exactly: There is no orbital magnetism apart from the quantum
fluctuations. Hence, $\mm_{\rm edge}$ characterizes those few (edge)
states which carry a finite current along the boundary, balancing the
bulk magnetization due to their large positive magnetic moments.

The edge magnetization is well-defined in the exterior as well.
There, it is negative with the mean like \eref{eq:mmedgesm} but for a
minus sign in front of the area term.  This suggests to define the
edge state density as $\dedge(\nu)= \pm\mm_{\rm edge}(\nu)/\nu$, with
the lower sign for the exterior problem.  The corresponding weights
\begin{equation}
  \label{eq:wndef}
  w_\wasn=\pm \frac{b^2}{\nu_\wasn}\, \frac{\rmd \nu_\wasn}{\rmd b^2}
  =
  \pm\frac{1}{\nu}
  \left(
    \frac{\bra{\psi_\wasn}\rvec\times\vvec\ket{\psi_\wasn}}{\omega_{\rm c}b^2}
    +\nu
  \right)
  >0
\end{equation}
are easily obtained as the derivative of the eigenenergies taken at
fixed $\rho$.  This definition satisfies the conditions formulated
above.  In particular, the weights are
exponentially small for bulk states 
since the Landau energies $\nu=N+\oh$,
$N\in\Natural_0$, are independent of $b$ \cite{Hornberger01}.  

The  semiclassical edge state density is
derived by inserting the
trace formula for $\N(\nu)$ \cite{Gutzwiller71,PAKGEC94} into
\eref{eq:mmedgedef}.  One obtains a sum over all skipping periodic
orbits for the fluctuating part $\dosc_{\rm edge}=\dedge-\dedgesm$. It
differs from the semiclassical expression of the unweighted spectral
density only by
a  factor
\begin{equation}
  \label{eq:wpodef}
   w_\po
   =
    \frac{ 2 \Area_\po\pm\rho\Len_\po}{\rho\Len_\po}
\end{equation}
attributed individually to each periodic orbit contribution.
This {\em classical weight}
 is determined by the area $\Area_\po$ enclosed by the trajectory $\po$
and by its length $\Len_\po$.
The  weights
approach zero as the skipping orbits are further detached from the
boundary.  
Hence,
the classical weights  \eref{eq:wpodef}
 smoothly
suppress
the bulk contributions to the semiclassical spectral density.

\begin{figure}
  \begin{center}%
    \includegraphics[width=.86\linewidth]{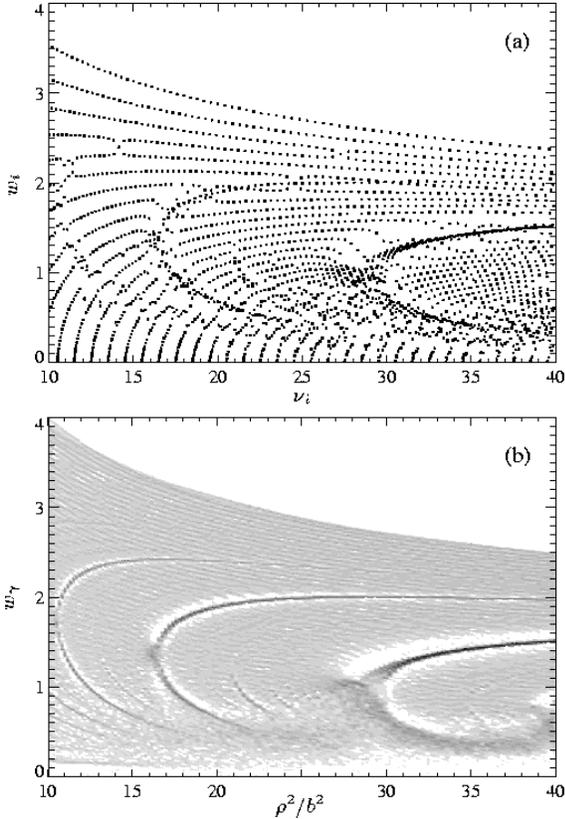}
  \end{center}%
  \caption{%
    (a) Weighted edge state spectrum
    compared to (b) the phase space distribution of the classical
    weights for the interior ellipse.  The quantum weights
    \eref{eq:wndef} segregate the edge states from the bulk and
    mimic the structures in the distribution of
    classical weights \eref{eq:wpodef}.
    }
\label{fig:qmclswt}
\end{figure}

Fig.~\ref{fig:qmclswt} shows (a) the quantum weights $w_\wasn$ against
their energies $\nu_\wasn$ \cite{HS00a}
and (b) the phase space distribution of the classical
weights (for the interior ellipse billiard with area $\Area=\pi$ and
eccentricity $0.8$  at $b=0.1$; 
the shade in (b) gives the 
probability measure for finding a
trajectory with weight $w_\po$.) 
In Fig.~\ref{fig:qmclswt}(a) one observes
 how the $w_\wasn$ distinguish the relevant edge states
from the bulk.  Characterized by vanishingly small $w_\wasn$, the bulk
states accumulate at the Landau levels, $\nu=N+\oh$, with sequences of
transitional states connecting to the large edge weights.  The latter
distribute in structures
reproduced by
the classical weights \eref{eq:wpodef} (which are
due to bifurcating regular islands in phase space).  In the exterior,
the segregation into edge and bulk is even more distinct (not shown).
\cite{note2}

In order to unravel the relation  between the interior and the
exterior spectra we consider the cross-correlator
\begin{eqnarray}
  \label{eq:corrdef}
  C(\nu_0) &=&
  \int\!\!\!\int
  \dosc^{\rm (int)}_{\rm edge}(\nu;b^{ 2})\,
  \dosc^{\rm (ext)}_{\rm edge}(\nu;b^{ 2})\,
  h\!\left(\frac{b^{ 2}- b_0^2}{b_0^2}\right)
   \frac{\rmd b^{ 2}}{b_0^2}
\nnn
&&\quad\times
 \,  g(\nu-\nu_0)
   \, \rmd \nu
\CO
\end{eqnarray}
defined for fixed reference energy $\nu_0$ and magnetic length $b_0$.
The functions $g$ and $h$ are normalised Gaussians.  $g$ regularises
the integral over the pair distribution $\dosc^{\rm (int)}_{\rm
  edge}(\nu;b^{ 2})\, \dosc^{\rm (ext)}_{\rm edge}(\nu;b^{ 2})$ and  $h$
restricts the $b^2$ integration to a range where a linear expansion of
$\nu_n(b^2)$ near $b_0^{ 2}$ is allowed. Then, $C(\nu_0)$ can be
written as a double sum over the interior and the exterior edge
spectra obtained at one fixed value $b_0$ of the magnetic length:
\begin{eqnarray}
  \label{eq:qmcorr}
  C(\nu_0)&=&
  \sum_{i,j=1}^\infty
 \!\frac{w_i w_j'}{w_i+w_j'}\,
  g\!\left(
    \frac{  \frac{\nu_i-\nu_0}{w_i}-\frac{\nu_0 - \nu_j' }{w_j'}       }
    { \frac{1}{w_i}+\frac{1}{w_j'} }
  \right)
  h\bigg(
    \frac{\nu_i-\nu_j'}{w_i+w_j'}
  \bigg)
\nnn
&&- C_{\rm background}
\end{eqnarray}
The primes label the exterior energies and weights. 
 Since the width
of $g$ is taken small, $\sigma_g\ll (\dedgesm)^{-1}$, only those pairs of
interior and exterior energies contribute 
whose distances to $\nu_0$ {\em scaled} by the the respective quantum
weights are approximately equal.  The prefactor in \eref{eq:qmcorr}
ensures that  only pairs of edge energies contribute.

When  evaluating $C(\nu_0)$ semiclassically, the integration over
$b^{ 2}$ in \eref{eq:corrdef} selects those pairs  in the sum
over
interior and exterior orbits $\po$ and $\po'$ which satisfy
$w_\po\Len_\po \simeq w_{\po'} \Len_{\po'}$,
a relation fulfilled by the {\em dual} pairs.
Restricting the summation to the
latter
is tantamount to the ``diagonal approximation'' \cite{Berry85}. 
The actions complement each other to $2\pi\nu_0 n_\po$ where
$2\pi\nu_0$ is the scaled action of a cyclotron orbit, and $n_\po$
is the number of reflections.
We obtain
\begin{eqnarray}
  \label{eq:sccor3}
  C(\nu_0)
    &=&
    \sum_{n=n_{\rm min}}^\infty
    f(n) \, \hat{g}(n)\,
    \cos(2\pi n(\nu_0-\tfrac{1}{2}))
\end{eqnarray}
with  $\hat{g}$ the Fourier transform of $g$.
Here, $n_{min}$ is the minimal number
of reflections needed for a periodic orbit at given $\rho$ and
\begin{equation}
  \label{eq:fn}
  f(n) =
  \frac{2}{\pi}
  \sum_{\po\in \Gamma_n }
   a_\po^2 w_\po^2 
\end{equation}
a sum over 
the set $\Gamma_n$ of dual
orbits with $n$ reflections.
It involves the classical weights $w_\po$ \eref{eq:wpodef} and the stability
amplitudes $a_\po$ \cite{Gutzwiller71} of the unweighted spectral density.

From its semiclassical form  \eref{eq:sccor3} the correlator
is expected to be appreciably
different from zero only at energies
where the cosine terms are stationary. Hence, $C(\nu_0)$ must
exhibit  peaks
at $\nu_0=N+\oh$. Its Fourier transform, on the other hand,
should
be peaked at the integer values starting from $n_{\rm min}$.

\begin{figure}[p]%
  \begin{center}%
    \includegraphics[width=0.9\linewidth]{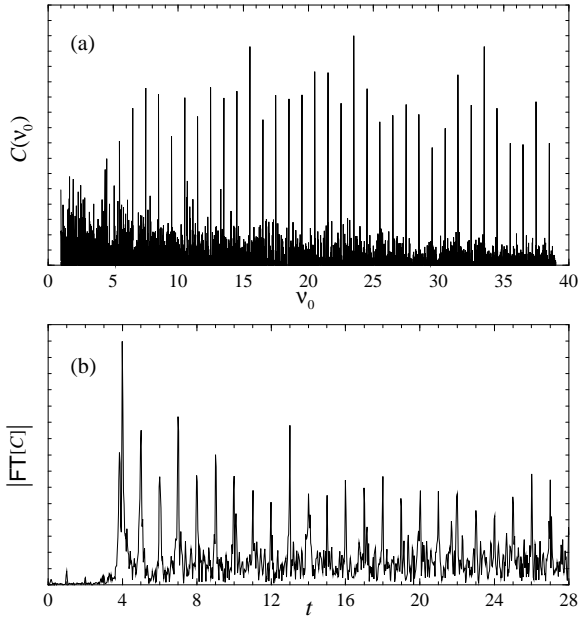}
  \end{center}%
    \caption{%
      (a) Cross-correlation function \eref{eq:qmcorr}
      for the      elliptic billiard.
      The pronounced spikes
      indicate the existence of pairwise
      correlations between interior and exterior edge states.
      (b)  Fourier transform of
      \eref{eq:qmcorr}.
      The peaks at integers starting from $4$ prove the
      classical origin of the correlations
      ($\sigma_g=0.001$).
  }
\label{fig:Cellipse}%
\end{figure}

A numerical verification of these predictions is presented in
Fig.~\ref{fig:Cellipse} for the spectra of the ellipse billiard at
$b_0=0.1$ \cite{HS00a}.
In Fig.~\ref{fig:Cellipse}(a)
one observes that the cross correlation function 
is strongly fluctuating but
displays pronounced spikes at the expected energies.
They are a clear signature of a correlation between the interior and
exterior {edge} states.
The  peaks in the Fourier transform,
Fig.~\ref{fig:Cellipse}(b),  are positioned at integer values of $t$ which
start at $n_{\rm min}=4$, as expected for the (desymmetrized)
ellipse. They  expose clearly the classical duality as the origin of
the cross-correlations.

As a test we restricted the sum \eref{eq:qmcorr} to pairs taken from
the two different symmetry classes of the ellipse.  This erases the
peaks in $C(\nu_0)$ as one expects semiclassically.  On the other
hand, removing the bulk states by imposing a threshold on the $w_\wasn$
does not change Fig. \ref{fig:Cellipse}.

The spikes of $C(\nu_0)$
imply the existence of a  {\em
  pairwise} relation between the interior and exterior edge states:
For each correlated pair of interior and exterior edge energies,
$\nu_i$ and $\nu_j'$, there exists a Landau level $\nu_0=N+\oh$ such that the
distances -- scaled individually by the 
quantum weights $w_i$
and $w_j'$
-- are approximately equal,
\begin{equation}
  \label{eq:pairrel}
  \frac{\nu_i-(N-\tfrac{1}{2})}{w_i}
  \simeq
  \frac{(N-\tfrac{1}{2})-\nu_{j}'}{w_{j}'}
\PO
\end{equation}
This follows immediately from \eref{eq:qmcorr}, where we took the width
of $g$ to be small on the quantum scale.
The relation \eref{eq:pairrel} shows that the interior and exterior
edge spectra are intimately connected in the semiclassical limit.
Note in particular the vital
role played by the quantum weights \eref{eq:wndef}
without which the correlations would not be observable.

Equation \eref{eq:pairrel} allows  to spot single pairs of correlated
states in the spectrum, and Fig.~\ref{fig:cpair}(b) gives an
outstanding illustration (with the shade proportional to $|\psi|^2$):
The wave functions are clearly localized along the {stable} dual
periodic orbits drawn in Fig.~\ref{fig:cpair}(a).  Although
the respective energies are separated by 20 mean edge state spacings,
$\nu_i=31.42554$, $\nu_j'=31.61696$, the difference between the two
sides of (\ref {eq:pairrel}) with $w_i=17.23/\nu_i$,
$w_j'=26.76/\nu_j'$, and $N=31$ is approximately one tenth of the mean
level spacing scaled by the mean weight.  The correlation of pairs of
{\em chaotic} wave functions cannot be verified as easily by visual
inspection, but they exhibit a large overlap of their normal
derivatives at the boundary.

Let us finally emphasize that the correlations between edge states of
the interior and the exterior do not permit us to derive one spectrum
from the other -- even in the semiclassical limit -- since the
respective Landau level $N$ and the complementary quantum weights are
not known a priori.  Nonetheless, they quantify a deep interconnection
of the spectra which is generated by the classical duality, as
initially conjectured from Fig. \ref{fig:cpair}.
Similar cross-correlations may be expected between  non-magnetic
systems as well, e.g. in  complementary billiards on the sphere.

We thank A. Buchleitner for helpful comments on the manuscript and B.
Gutkin for helpful discussions.  The work was supported
by the Minerva Center for Nonlinear Physics.


\begin{thebibliography}{10}
\bibitem{note2}
As shown in \cite{HS01a}, the edge state density may be defined alternatively
  by varying the boundary condition. The corresponding quantum and classical
  weights differ from \eref{eq:wndef} and \eref{eq:wpodef} but give rise to the
  same relation \eref{eq:pairrel}. Hence, the ratio between the quantum weights
  of corresponding states is semiclassically the same for both definitions.
\end{thebibliography}
\end{document}